\documentclass{llncs}

\usepackage{xspace}
\usepackage{graphicx}
\usepackage{amsmath}
\usepackage{amsfonts}
\usepackage{url}
\usepackage{array}
\usepackage{colortbl}

\newcommand{\QL}{\ensuremath{Q\!L}\xspace}

\newcommand{\CL}{\ensuremath{C\!L}\xspace}

\newcommand{\MA}{\ensuremath{M\!A}\xspace}

\newcommand{\cert}{\mathop{\ensuremath{cert}}}

\newcommand{\cQ}{\mathop{\ensuremath{cQ}}}

\newcommand{\bQ}{\mathop{\ensuremath{bQ}}}

\newcommand{\Ind}{\mathop{\ensuremath{I\!nd}}}

\newcommand{\tbox}[1]{\mathtt{#1}}
\newcommand{\abox}[1]{\mathtt{#1}}

\title{%
  Evaluating Modelling Approaches for Medical Image Annotations
}
\author{Jasmin Opitz, Bijan Parsia, Ulrike Sattler}
\institute{The University of Manchester\\
 \email{$\{$opitzj|bparsia|sattler$\}$@cs.manchester.ac.uk}\\[3ex]
}

\begin{document}
\maketitle

\begin{abstract}
Information system designers face many challenges w.r.t. selecting appropriate semantic technologies and deciding on a modelling approach for their system. However, there is no clear methodology yet to evaluate ``semantically enriched'' information systems. In this paper we present a case study on different modelling approaches for annotating medical images and introduce a conceptual framework that can be used to analyse the fitness of information systems and help designers to spot the strengths and weaknesses of various modelling approaches as well as managing trade-offs between modelling effort and their potential benefits.

\end{abstract}

\section{Introduction}
Information systems can have very different shapes and variants and designing such systems involves taking important modelling decisions to optimise information retrieval. The different dimensions to be taken into account are performance, maintainability and usability of the system. Some research has been carried out to evaluate information systems w.r.t. these dimensions \cite{Chandler1976,Harmon1999,Morse2002}, however, none of these approaches analyses the quality of such a system w.r.t. its queriability, i.e. how easy or comfortable it is for a user to formulate queries. Since information retrieval is the main purpose of an information system, we believe that the queriability and therefore the difficulty of assessing the information is a crucial point.

Recently, there has been a lot of discussion about ``semantically enriched'' information systems, especially about using ontologies for modelling data. Ontologies have the potential of modelling information in a way that they can capture the ``meaning'' of the content by using expressive knowledge representation formalisms (such as Description Logics \cite{Baader2003}) and therefore achieve good information retrieval results. However, ontology-based information systems can come in many different fashions. It has to be decided how expressive the conceptual schema is, whether it is possible to use an off-the-shelf schema or one that is tailored to the application. Furthermore, it is crucial that the schema and the data are well-suited in order to enable good queriability and retrieval performance. Depending on the design decisions, more or less modelling effort is involved. Information system designers need a methodology that helps them to understand the benefits and trade-offs of various approaches and to make an informed decision about the ``optimal'' modelling approach for their needs.\\

In this paper we present a case study on medical image annotations. Medical images are usually stored in large databases or file systems along with various information about the images, such as radiologists reports (usually formulated in natural language). Although it is possible to retrieve images based on these descriptions with full-text search and keyword queries, the results are prone to have a low recall and precision \cite{Opitz2009}. There is a wide range of alternative modelling approaches. The relevant terms in the natural language descriptions could be mapped to an underlying schema, such as a thesaurus or to an ontology. The more expressive the schema, the more ``meaning'' can be potentially modelled in the image annotations. This might result in better retrieval performance but also in more modelling effort while creating the annotations, e.g. depending on whether or not the information extraction can be done automatically or has to be done manually.

There is a trade-off between how much effort is involved in the design of the system and the creation of meaningful image annotations and how this leads to better queriability and better quality of the retrieval results. In order to understand this trade-off and to analyse and compare different approaches for modelling medical image annotations we will use a conceptual framework \cite{Opitz2010} that has been designed for the evaluation of information systems with a particular focus on queriability. We will outline five modelling approaches for medical image annotations that range from a simple text-based information system to full-fledged ontology-based information systems based on an established medical ontology, namely SNOMED CT.\footnote{SNOMED CT: \url{http://www.ihtsdo.org/snomed-ct/}} Applying the framework to these modelling approaches will highlight their strengths and weaknesses and the framework's measurements will allow us to compare the modelling approaches with each other. This case study is one of several studies that we are carrying out to evaluate and refine the framework.

The remainder of the paper is organised as follows. In Section \ref{mas} we outline several possible modelling approaches for medical image annotations. Section \ref{framework} introduces the Evaluation Framework and how it can be applied to measure the fitness of a modelling approach. In Section \ref{imageAnnotations} we apply the framework to the particular modelling approaches that we outlined for medical image annotations. We measure and compare their fitness and queriability and discuss the weak and strong points of each modelling approach. Section \ref{conclusion} concludes the paper.

\section{Modelling Approaches for a Medical Image Information System} \label{mas}
An information system for medical image annotations can be designed in many different ways. In the case study described in this paper we distinguish between three general categories of modelling approaches. A simple text-based information system, a slightly more advances thesaurus-based information system and an ontology-based information system. For the latter we distinguish between three different ways of designing an ontology-based information system, differing in the expressivity of the underlying schema and the annotations.

All of the five modelling approaches are based on the same data corpus. We used 42 publically available chest radiology images and their natural language (English) radiology reports from the web-based radiology database EURORAD.\footnote{EURORAD: \url{http://eurorad.org/}} The reports contain information about image type and modality, findings, body parts, diagnoses, etc. An early experiment on image retrieval with this data set has been published in \cite{Opitz2009}. The original data, i.e. the natural language reports, are processed differently for each of the five modelling approaches.

The various approaches described below are representative. Obviously, there are other alternatives or mixtures between those mentioned. Some of them could be partly improved with some customising. The selection in this paper represents distinct groups of modelling approaches that we chose deliberately to highlight their strengths and weaknesses.

\subsection{Text-based Modelling Approach}
In a text-based modelling approach ($\MA_1$) for our medical image information system the data collection consists merely of the unprocessed natural language descriptions. The queries are conjunctions of natural language keywords and the query results are obtained by carrying out a full-text search over the image descriptions.

This method involves very little modelling effort and simple querying (like in an internet search engine). However, the approach is not very powerful: neither the queries nor the underlying data can identify synonyms, homonyms, acronyms, spelling mistakes or capture taxonomical or relational information. Therefore, a text-based approach is not suitable for capturing the full semantics of the image descriptions and most likely leads to low recall and precision of the retrieval results.

\subsection{Thesaurus and Text Mining based Modelling Approach}\label{thesaurusBased}
For this modelling approach ($\MA_2$) we processed the natural language descriptions with publically available, off-the-shelf text mining tools. We used GENIA tagger\footnote{GENIA tagger: \url{http://www-tsujii.is.s.u-tokyo.ac.jp/GENIA/tagger/}} to extract noun phrases, verb phrases and adjective phrases from the textual descriptions and processed these with MetaMap\footnote{MetaMap: \url{http://metamap.nlm.nih.gov/}} in order to map them to SNOMED CT classes via UMLS.\footnote{UMLS: \url{http://www.nlm.nih.gov/research/umls/}}
For each image an annotation file was created containing a list of SNOMED CT classes that reflect the relevant terms extracted from the text. Although SNOMED CT is strictly speaking an ontology that can be expressed in OWL and conforms to the OWL 2 profile OWL EL \cite{owl-spec}, we merely use it in the sense of a thesaurus for this modelling approach, i.e. each image is ``tagged'' with a list of SNOMED CT classes and we only make use of the synonyms and the taxonomical information that are contained in the ontology and do not capture any relational information.

Again, the queries are expressed in natural language and are processed in the same way and with the same tools as the image descriptions. Each query is transformed to a list of SNOMED CT classes and matched against the image annotations.

The advantages of a thesaurus-based modelling approach compared to a purely text-based approach are that recall and precision can be increased due to the fact that a thesaurus can recognise synonyms to a certain extent and contains taxonomical information so that image annotations that contain subclasses of the query terms are retrieved as well \cite{Krauthammer2004}. The additional modelling effort required to achieve these benefits is limited to incorporating an off-the-shelf thesaurus and processing the textual descriptions automatically with off-the-shelf text mining tools. On the other hand, this approach does not capture relational information and the tools do not map all the available information perfectly, i.e. the translation process is error prone (e.g. mapping to wrong class or no mapping at all, inability to recognise acronyms).

\subsection{Ontology-based Modelling Approaches}
All of the three ontology-based modelling approaches we discuss in this paper ($\MA_3$, $\MA_4$ and $\MA_5$) involve translating the textual image descriptions manually to Abox assertions to different schemas (SNOMED CT Tboxes). The queries are formulated as OWL class expressions.

We distinguish between three variants of ontology-based annotations:
\begin{itemize}
\item $\MA_3$: image annotation Abox ($\abox{A}_1$) with class assertions to a SNOMED CT Tbox ($\tbox{T}_1$)
\item $\MA_4$: image annotation Abox ($\abox{A}_2$) with class and role assertions to a SNOMED CT Tbox ($\tbox{T}_1$)
\item $\MA_5$: image annotation Abox ($\abox{A}_3$) with class and role assertions to a SNOMED CT Tbox ($\tbox{T}_2$) that contains some additional ``image-annotation-specific'' roles
\end{itemize}

$\MA_5$ uses a slightly different Tbox than $\MA_3$ and $\MA_4$ in the sense that we created an additional set of roles and a role hierarchy in order to bypass the SNOMED CT specific role groups.  An example of a disease in SNOMED CT that is defined using role groups is \emph{NeoplasmOfLung}. The concept is defined as follows:\footnote{To improve readability, we use slightly abbreviated class names and DL syntax.}

\begin{small}
$$\begin{array}{rl}
\mathrm{NeoplasmOfLung}\equiv &\mathrm{DisorderOfLung}\sqcap~ \\
&\exists \mathrm{roleGroup(}
\begin{array}[t]{l}
  \exists \mathrm{AssociatedMorphology.Neoplasm}\sqcap~ \\
                               \exists \mathrm{FindingSite.LungStructure)} 
\end{array}
\end{array}$$
\end{small}

For $\MA_5$, we introduced three additional roles: \emph{shows}, \emph{hasFinding} and \emph{hasLocation} and defined the following role hierarchy:

\begin{small}
$$
\begin{array}[t]{rcl}
\mathrm{roleGroup} ~ \circ ~ \mathrm{AssociatedMorphology} &\sqsubseteq& \mathrm{hasFinding}\\
\mathrm{roleGroup} ~ \circ ~ \mathrm{FindingSite} &\sqsubseteq &\mathrm{hasLocation}\\
\mathrm{shows} ~ \circ ~ \mathrm{hasFinding}& \sqsubseteq& \mathrm{shows}\\
\mathrm{shows} ~ \circ ~ \mathrm{hasLocation}& \sqsubseteq& \mathrm{shows}
\end{array}$$
\end{small}

If we want to find all images that show neoplasms in $\MA_5$, we can formulate a simple OWL class expression query like \begin{small}$~\mathrm{Image} \sqcap \exists \mathrm{shows.Neoplasm}~$\end{small} and would retrieve images labelled with \begin{small}$~\mathrm{Image}\sqcap \exists \mathrm{roleGroup.} \exists \mathrm{AssociatedMorphology.NeoplasmOfLung}~$\end{small} without having to use the complicated role group construct in the query.

Furthermore, we introduced the role \emph{derivingFrom} in order to indicate a causal relationship between two findings, e.g. a metastasis and a primary tumour.\\

Similar to the thesaurus-based modelling approach (see Section \ref{thesaurusBased}) we use an off-the-shelf schema. By using an underlying ontology we can take advantage of all the benefits that a thesaurus-based modelling approach involves (e.g. synonyms and taxonomical relations between classes). $\MA_4$ and $\MA_5$ additionally capture relational information and are therefore better suited to capture the actual meaning of the image descriptions and increase recall and precision of the results compared to the other modelling approaches.

On the other hand, all of these modelling approaches involve a high modelling effort due to the manual translation of natural language to ontology-based annotations. This requires both domain knowledge as well as knowledge of OWL. Furthermore, the benefits that these approaches potentially implicate can only be taken advantage of if the schema, data and queries are well-suited.

\section{Information System Evaluation Framework}\label{framework}
We start by formalising the relevant components of a (semantically enriched) information system for which we are then going to evaluate and compare the different modelling approaches. We will use the term ``modelling approach''  to describe the whole system consisting of data, schema, (an abstraction of) queries, and a query language. A more detailed description of this framework can be found in \cite{Opitz2010}.

\subsection{A Modelling Approach}

A \emph{modelling approach} $\MA = (S,D,R, \QL)$ consists of

\begin{itemize}
\item a schema $S$: a finite description of the semantics of the data, e.g. a database schema, a logic program, or the Tbox of an ontology, which can be empty. 
\item the data $D$: e.g. tables and rows in a relational database, ground facts, or ontology Abox assertions. 
\item a set of information requests $R$: each $r\in R$ represents the answer to a query of $D$, and is given as a set (of tuples) over $D$. Ideally, $R$ should be representative for the queries to be answered by the information system to be built.
\item a query language \QL: e.g. SQL, (union of) conjunctive queries, OWL class expressions. 
\end{itemize}

An information request asks for tuples of the given data that are relevant for the user. The request needs to be distinguished from the actual query, which is a specific manifestation of the information request formulated in \QL. An information request $r$ can correspond to 0, 1 or more queries in a given query language. The former is the case if there are no queries in \QL  whose answers would be exactly the tuples in $r$ when asked over $S$ and $D$, i.e., if \QL is unable to express the information request over the given schema and data. In the case that there are one or more queries, some of them might be more easily expressible than others.

The only assumptions we make is that the query language \QL comes with a semantics that identifies, for a given query $q$ of arity $n$ in \QL, data $D$, and schema $S$, the set of \emph{certain answers} \cite{Calvanese1998}. More precisely, we assume the existence of an entailment relation $\models$, and use $\Ind(D)$ for the set of individuals or constants in $D$ to define  $\cert(\cdot)$ as follows:  
\begin{small}
$$\cert(q,S,D) = \{ \vec w \in \Ind(D)^n \mid S \cup D \models q(\vec w) \} 
.$$
\vspace{-1cm}
\end{small}

\subsection{Measuring the Fitness of a Modelling Approach}
The basic characteristic we want to evaluate is the \emph{fitness} of a modelling approach, i.e. how well the schema and the data are suited to enable the formulation of ``fit'' queries for answering the given information requests. The fitness of a modelling approach can be determined by analysing the syntactic, semantic and/or cognitive complexity of the queries that correspond to the information requests and depends on the \emph{fitness function}.

\subsubsection{The Fitness Function}
Different queries that correspond to an information request can vary in length and be more or less complex, e.g. in terms of using relations and constructors such as conjunctions, disjunctions, etc. They can also be more or less difficult to understand from a cognitive perspective. For example, a human user might find a query that uses terms that are actual words (in the sense that they exist in a domain expert's dictionary) easier to understand than one that uses anonymous identifiers.  The purpose of the fitness function is to capture this complexity. 

The framework is parameterised with a \emph{fitness function} $f$ that associates each query $q$ in \QL with some value $f(q)$ that is intended to capture its fitness. We only require that $f$ maps \QL into a  totally ordered set $(M,<)$, e.g. $\mathbb{R}$ or $\mathbb{N}^4$, which we call the query's \emph{fitness value}.  Obvious examples of fitness functions are (i) a query's length, (ii) a query's length combined with the number of constructors involved, either via some (weighted) summation or into a vector, or (iii) a query's length combined with the number of terms not to be found in a domain expert's diactionary, or any combinations or extensions of these. 

The smaller the fitness value, the ``better'' the query. We read $f(q) < f(q')$ as $q$ being ``better'' or ``fitter'' than $q'$.  The framework evaluates the ``best queries'' for an information request, e.g., the shortest and least complex queries. The fitness function induces a partial order on the queries.

\subsubsection{The Query Space}
Each information request $r\in R$ has an associated query space: first, we define \emph{correct queries} $\cQ(r,S,D)$  as those that answer exactly an information request $r$ over $S$ and $D$:

\begin{small}
$$\cQ(r,S,D) = \{ q ~ \mid  q \text{ is a } \QL \text{ query and } \cert(q, S, D) = r(D) \}.$$
\end{small}

Next, we define \emph{best queries} $\bQ(r,S,D,f)$ as those correct queries whose fitness is maximal. Clearly, best queries depend on how we measure fitness, and thus on the fitness function $f$: 

\begin{small}
$$\bQ(r,S,D,f) =  \{ q  \in \cQ(r,S,D) \mid
\begin{array}[t]{l}
  \text{ there is no }q'\in \cQ(r,S,D) \colon \\ \multicolumn{1}{r}{f(q') < f(q) \} .}
\end{array}
$$
\end{small}

Since the $\bQ(\cdot)$ are the ``fittest'' queries among the correct queries, any two queries in  $\bQ(\cdot)$ are equally fit, and we can abbreviate their fitness as follows: for $ f(q_i) = f(q_j)$,  we set $f(\{q_1,...,q_k\})$ to be $ f(q_1)$. 

\subsection{Using the Evaluation Framework to Compare Modelling Approaches}
If we want to compare several modelling approaches, we can compare, for each information request $r_i\in R$ and each of the modelling approaches, the fitness of the best queries. This can unveil the strengths and weaknesses of the information system to the system designer. For example, if there are information requests for which the set of correct queries is empty, then $f(\bQ(r,S,D,f))$ is prohibitively bad. To overcome this, we can then decide whether to select a different, more powerful query language or to change the schema or the way the data is modelled---or whether perhaps that particular information request is of too little importance for such a change. The measurements can also help to point out where the trade-offs between modelling effort and benefits in terms of easier query answering are. For example, considering an ontology-based modelling approach, whether more modelling effort for a more expressive Tbox would be justified for the sake of simpler queries.

Applying the framework to \emph{one} modelling approach $\MA = (S,D,R,\QL)$ reveals for each $r$, the fitness value of the best queries: $f(bQ_j)$. In particular, it will identify information requests for which it is hard to specify a query in \QL and those for which this is impossible. When comparing different modelling approaches we can compare the point-to-point fitness for each information request.

\section{Applying the Framework to a Medical Image Annotation System}\label{imageAnnotations}
We will now demonstrate how the evaluation framework can be applied to the various modelling approaches we sketched in Section \ref{mas}. The goal of this evaluation is to find out how to model the information in order to get optimal retrieval results, how much effort is involved in this ``optimal'' modelling and how fit the modelling approaches are w.r.t. queriability.

\subsection{Information Requests}
A set of representative information requests $R$ is derived from the content of the original, natural language image descriptions: image types and modalities, clinical findings, complex findings (e.g. involving locations) and combinations of the former.

\begin{itemize}
\item $r_1$: involves one clinical finding: ``All images that show neoplasms.''
\item $r_2$: involves two concepts, an image type and an image projection: ``All X-ray images with lateral projection.''
\item $r_3$: involves a clinical finding combined with a qualifier value: ``All images that show left-sided pleural effusions.''
\item $r_4$: involves a clinical finding combined with a body structure: ``All images that show soft tissue masses in the pleural membrane.''
\item $r_5$: involves a causal relationship between two findings: ``All images with metastases deriving from a carcinoma.''
\end{itemize}

The information requests are extensional, i.e. they are mapped to a set of answers from the data space. Above, we listed English descriptions of $r_1$ to $r_6$ for better understanding.

\subsection{Modelling Approaches}
We formalise the modelling approaches that we introduced in Section \ref{mas}:
\begin{itemize}
\item $\MA_1 = (\emptyset$, text, $R$, keywords$)$
\item $\MA_2 = (\tbox{T}_1,$ concept list, $R$, keywords$)$
\item $\MA_3 = (\tbox{T}_1,\abox{A}_1,R,\CL)$
\item $\MA_4 = (\tbox{T}_1,\abox{A}_2,R,\CL)$
\item $\MA_5 = (\tbox{T}_2,\abox{A}_3,R,\CL)$
\end {itemize}

Note that all representation of the data are derived from the same data set, i.e. the original natural language image descriptions. In the various modelling approaches we tried to extract information from this corpus in different ways, some of which capture more ``meaning'' than others.

\subsection{Results}
Table 1 and Table 2 show for each of the above mentioned information requests and modelling approaches both the best query and its fitness vector of the form $m=(x,y,z)$ where $x$ is the length of the query, $y$ is the number of distinct constructors and $z$ indicates the nesting depth. For better readability of the queries we abbreviated some of SNOMED CT's role names: AM stands for \emph{AssociatedMorphology}, FS stands for \emph{FindingSite}, QV stands for \emph{QualifierValue} and rG stands for \emph{roleGroup}.

\begin{table}
\centering
\begin{scriptsize}
\begin{tabular}{|c|p{3cm}|p{3.5cm}|p{4.5cm}|}
\hline
\rule{0pt}{2pt}
$r_j$ & $\MA_1$ & $\MA_2$ & $\MA_3$\\
\hline
$r_1$ & \cellcolor[gray]{0.85} neoplasm & Neoplasm & $\begin{array}[t]{l} \mathrm{Image} \sqcap  ~\\ ~\exists \mathrm{shows.Neoplasm} \end{array}$\\
&\cellcolor[gray]{0.85}&&\\
& \cellcolor[gray]{0.85}$m_{11}=(1,0,0)$ & $m_{12}=(1,0,0)$ & $m_{13}=(3,2,1)$\\
\hline
$r_2$ & \cellcolor[gray]{0.85}$\begin{array}[t]{l} \mathrm{xray} \sqcap ~\\ \mathrm{lateral} \end{array}$ & $\begin{array}[t]{l} \mathrm{Xray} \sqcap ~\\ \mathrm{LateralProjection} \end{array}$  & $ \begin{array}[t]{l} \mathrm{Image} \sqcap  ~\\ ~\exists \mathrm{shows.Xray} \sqcap  ~\\ ~\exists \mathrm{shows.LateralProjection} \end{array}$\\
&\cellcolor[gray]{0.85}&&\\
& \cellcolor[gray]{0.85}$m_{21}=(2,1,0)$ & $m_{22}=(2,1,0)$ & $m_{23}=(5,2,1)$\\
\hline
$r_3$ & \cellcolor[gray]{0.85}$\begin{array}[t]{l} \mathrm{pleural ~ effusion} \sqcap ~\\ \mathrm{left} \end{array}$ & \cellcolor[gray]{0.85}$\begin{array}[t]{l} \mathrm{PleuralEffusion} \sqcap ~\\ \mathrm{LeftSided} \end{array}$ & \cellcolor[gray]{0.85}$\begin{array}[t]{l} \mathrm{Image} \sqcap  ~\\ ~\exists \mathrm{shows.PleuralEffusion} \sqcap  ~\\  ~\exists \mathrm{shows.LeftSided} \end{array}$\\
&\cellcolor[gray]{0.85}&\cellcolor[gray]{0.85}&\cellcolor[gray]{0.85}\\
& \cellcolor[gray]{0.85}$m_{31}=(2,1,0)$ & \cellcolor[gray]{0.85}$m_{32}=(2,1,0)$ & \cellcolor[gray]{0.85}$m_{33}=(5,2,1)$\\
\hline
$r_4$ & \cellcolor[gray]{0.85}$\begin{array}[t]{l} \mathrm{soft ~ tissue ~ mass} \sqcap ~\\ \mathrm{pleural ~ membrane} \end{array}$ & \cellcolor[gray]{0.85}$\begin{array}[t]{l} \mathrm{SoftTissueMass} \sqcap ~\\ \mathrm{PleuralMembrane} \end{array}$ & \cellcolor[gray]{0.85}$\begin{array}[t]{l} \mathrm{Image} \sqcap  ~\\ ~\exists \mathrm{shows.SoftTissueMass} \sqcap  ~\\  ~\exists \mathrm{shows.PleuralMembrane} \end{array}$\\
&\cellcolor[gray]{0.85}&\cellcolor[gray]{0.85}&\cellcolor[gray]{0.85}\\
& \cellcolor[gray]{0.85}$m_{41}=(2,1,0)$ & \cellcolor[gray]{0.85}$m_{42}=(2,1,0)$ & \cellcolor[gray]{0.85}$m_{43}=(5,2,1)$\\
\hline
$r_5$ & \cellcolor[gray]{0.85}$\begin{array}[t]{l} \mathrm{metastasis} \sqcap ~\\ \mathrm{carcinoma} \end{array}$ & \cellcolor[gray]{0.85}$\begin{array}[t]{l} \mathrm{Metastasis} \sqcap ~\\ \mathrm{Carcinoma} \end{array}$ & \cellcolor[gray]{0.85}$\begin{array}[t]{l} \mathrm{Image} \sqcap  ~\\ ~\exists \mathrm{shows.Metastasis} \sqcap  ~\\  ~\exists \mathrm{shows.Carcinoma} \end{array}$\\
&\cellcolor[gray]{0.85}&\cellcolor[gray]{0.85}&\cellcolor[gray]{0.85}\\
& \cellcolor[gray]{0.85}$m_{51}=(2,1,0)$ & \cellcolor[gray]{0.85}$m_{52}=(2,1,0)$ & \cellcolor[gray]{0.85}$m_{53}=(5,2,1)$\\
\hline
\end{tabular}
\vspace{0.5cm}
\caption{Measurements for $\MA_1$, $\MA_2$ and $\MA_3$.}
\end{scriptsize}
\end{table}

\begin{table}
\centering
\begin{scriptsize}
\begin{tabular}{|c|p{6cm}|p{6cm}|}
\hline
\rule{0pt}{2pt}
$r_j$ & $\MA_4$ & $\MA_5$\\
\hline
$r_1$ & $\begin{array}[t]{l} \mathrm{Image} \sqcap \exists \mathrm{shows.(Disease} \sqcap  ~\\ ~ \exists \mathrm{rG.(}\exists \mathrm{AM.Neoplasm))}  \end{array}$ & $\begin{array}[t]{l} \mathrm{Image} \sqcap ~\\ ~\exists \mathrm{shows.Neoplasm} \end{array}$\\
&&\\
& $m_{14}=(6,2,3)$ & $m_{15}=(3,2,1)$\\
\hline
$r_2$ & $\begin{array}[t]{l}  \mathrm{Image} \sqcap   ~\\ \exists  \mathrm{hasImageType.Xray} \sqcap ~\\ \exists \mathrm{hasImageProjection.LateralProjection}\end{array}$ & $\begin{array}[t]{l}  \mathrm{Image} \sqcap   ~\\ \exists  \mathrm{hasImageType.Xray} \sqcap ~\\ \exists \mathrm{hasImageProjection.LateralProjection} \end{array}$\\
&&\\
& $m_{24}=(5,2,1)$ & $m_{25}=(5,2,1)$\\
\hline
$r_3$ & $\begin{array}[t]{l} \mathrm{Image} \sqcap  ~\\ ~\exists \mathrm{shows.(PleuralEffusion} \sqcap  ~\\  ~\exists \mathrm{QV.LeftSided)} \end{array}$ & $\begin{array}[t]{l} \mathrm{Image} \sqcap  ~\\ ~\exists \mathrm{shows.(PleuralEffusion} \sqcap  ~\\ ~\exists \mathrm{QV.LeftSided)} \end{array}$\\
&&\\
& $m_{34}=(5,2,2)$ & $m_{35}=(5,2,2)$\\
\hline
$r_4$ & $\begin{array}[t]{l}  \mathrm{Image} \sqcap  \exists \mathrm{shows.(Disease} \sqcap \exists \mathrm{rG.(} \\ ~\exists \mathrm{AM.SoftTissueMass} \sqcap  ~\\ ~\exists \mathrm{FS.PleuralMembrane))} \end{array}$ & $\begin{array}[t]{l}  \mathrm{Image} \sqcap   ~\\ ~\exists \mathrm{shows.(} \exists \mathrm{hasFinding.SoftTissueMass} \sqcap ~\\  ~\exists \mathrm{hasLocation.PleuralMembrane)} \end{array}$\\
&&\\
& $m_{44}=(8,2,3)$ & $m_{45}=(7,2,2)$\\
\hline
$r_5$ & \cellcolor[gray]{0.85}$\begin{array}[t]{l} \mathrm{Image} \sqcap ~\\ ~\exists \mathrm{shows.Metastasis} \sqcap  ~\\ ~\exists \mathrm{shows.(Disease} \sqcap \exists \mathrm{rG.(} \exists \mathrm{AM.Carcinoma))}  \end{array}$ & $\begin{array}[t]{l} \mathrm{Image} \sqcap   ~\\ ~\exists \mathrm{shows.(Metastasis} \sqcap ~\\  ~\exists \mathrm{derivingFrom.Carcinoma)} \end{array}$\\
&\cellcolor[gray]{0.85}&\\
& \cellcolor[gray]{0.85}$m_{54}=(8,2,3)$ & $m_{55}=(5,2,2)$\\
\hline
\end{tabular}
\vspace{0.5cm}
\caption{Measurements for $\MA_4$ and $\MA_5$.}
\end{scriptsize}
\end{table}

\subsection{Analysis}
The first observation we make is that not all of the queries noted in the tables return the correct answers to their corresponding information request. A query on a gray background indicates that it answers the information request only partially. In these cases the modelling approach is not expressive enough to capture the full meaning of the image description.
For example, the queries for $\MA_1$ are merely keywords, i.e. they cannot capture synonyms, homonyms etc. In $r_1$ we ask for image descriptions that contain the word ``neoplasm''. If we have a description that contains the synonym ``tumour'' instead, it will not be retrieved and the results therefore have a low recall. In $r_2$ we ask for ``Xray images with lateral projection.'' With the keywords ``xray'' and ``lateral'' we might also find images that show the ``lateral wall of the trachea'', but have a postero-anterior projection. In this case the query results have a low precision. We measured recall and precision in a preliminary study with the same data set and published the results in \cite{Opitz2009}.

The requests $r_3$ -- $r_5$ describe relational information, e.g. a finding in a particular location or a finding with a qualifier. Neither $\MA_1$ nor $\MA_2$ and $\MA_3$, however, can capture relational information. For example, if there is an image annotation that contains a ``soft tissue mass'' located in ``chest wall'' and a ``thickening'' of the ``pleural membrane'', it would be retrieved with any of the queries for $\MA_1$ -- $\MA_3$, although it is not an answer to $r_4$. Again, these queries lead to low precision.

The request $r_5$ expresses a causal relationship between two findings, i.e. ``metastases that derive from a carcinoma.'' Only $\MA_5$ is expressive enough to capture this information by using the \emph{derivingFrom} property that was added to SNOMED CT for this modelling approach.\\

The framework also helps us to compare the fitness of the modelling approaches. $\MA_1$ -- $\MA_5$ differ in expressivity, complexity and required modelling effort. $\MA_1$ is the least expressive, but also the least complex and involving comparatively little modelling effort. The textual image descriptions do not have to be processed in any way and the queries are just natural language keywords, i.e. the user does not have to know a particular query language. However, as stated above, recall and precision of the query results are lower than in the more expressive modelling approaches because $\MA_1$ does not take into account synonyms, homonyms, relational information etc.

The queries in $\MA_2$ are equally good as the ones for $\MA_1$. This approach does not involve a formal query language (we just used the keywords from the English representation of the respective information request which is then mapped to SNOMED CT terms automatically). Essentially, the user can use the same keywords as in $\MA_1$. Since $\MA_2$ can recognise synonyms and taxonomical relationships (which are defined in the thesaurus), its retrieval results are much better than those of $\MA_1$. It has to be noted however, that the tools we used to automatically map the text to concepts of the SNOMED CT ontology only know a finite number of synonyms for each concept. Therefore, natural language queries might lead to losses in recall. Losses in precision are also possible if homonyms are mapped to the wrong concept.

$\MA_3$ is very similar to $\MA_2$ in the sense that it merely lists SNOMED CT classes that were identified in the textual image descriptions. However, in $\MA_3$ the mapping has been done manually and not with the support of tools as in $\MA_2$. The manual mapping involved a considerably higher data modelling effort. Additionally, for $\MA_3$ the queries are formulated as OWL class expressions, which are more complex than natural language keywords. In summary, the queries in $\MA_2$ are slightly fitter than the ones in $\MA_3$ and the quality of the results for both modelling approaches is comparable.

$\MA_4$ and $\MA_5$ on the other hand are very expressive and capture the meaning of the original image descriptions very well and therefore lead to better recall and precision of the query results compared to the other modelling approaches. But that comes to the cost of a much higher modelling effort (due to the manual translation and the modelling of relational information) and higher complexity of the queries. An interesting observation is that a little more modelling effort on the schema of $\MA_5$ (i.e. adding some roles and role hierarchy axioms) greatly improves the queriability of this approach compared to the relatively similar approach $\MA_4$ without these axioms. Both in the image annotations and in the queries, cumbersome SNOMED CT specific constructs like \emph{roleGroups} can be avoided and lead to a better readability of the annotations and fitter queries and therefore increased queriability.\\

In general the evaluation framework helps us to understand:

\begin{itemize}
\item which are the weak spots of each modelling approach (e.g. low quality of query results in $\MA_1$, high complexity of the queries in $\MA_4$)
\item which are the strong spots of each modelling approach (e.g. $\MA_5$ captures the meaning particularly well and has high recall and precision)
\end{itemize}

With this knowledge the system designer can now assess the trade-offs between the advantages and disadvantages that these measurements indicate and the modelling effort that is required for each of the approaches. Based on this, informed decisions can be made. For example, that $\MA_2$ and $\MA_3$ are more or less similarly fit but $\MA_3$ requires significantly more effort due to the manual translation of the data. Therefore, $\MA_2$ might be preferred over $\MA_3$.

Another observation could be that if the precision of the results is not crucial, one might prefer a simpler modelling approach (e.g. $\MA_2$) that involves less modelling effort than the more expressive approaches but still leads to acceptable results. However, if recall and precision of the results are of high priority, one might want to invest some more effort and select $\MA_5$.

Furthermore, the evaluation makes salient how we could improve or customise a modelling approach in order to make it better or make use of its full potential. For example, modifying $\MA_4$ by adding some role axioms to the schema leads to a much fitter modelling approach $\MA_5$.

\section{Conclusion and Future Work}\label{conclusion}
We compared heterogeneous modelling approaches for an information system for medical image annotations. In order to evaluate the strengths and weaknesses of these approaches we used an information system evaluation framework that was designed to measure the fitness of modelling approaches with a particular focus on their queriability. The evaluation framework can make interesting characteristics of a modelling approach salient and therefore help information system designers to assess the trade-off between modelling effort and queriability and to make an informed decision about which modelling approach is most suitable for their needs.

Currently we are refining the evaluation framework so that it can be used for a wider range of scenarios and applications. We are extending the framework to measure not only the fitness of exact queries, but also such that answer information requests only partially. Furthermore, we incorporate measurements of false positives and false negatives in the query answers in order to combine the fitness of a query with retrieval performance measurements. Other possible extensions are e.g. measuring the flexibility of a system by evaluating in how many ways a user can formulate a fit query for an information request. Another adornment would be to combine the fitness values with measurements for query answering performance so that more information about the system can be taken into account. We are currently evaluating these extensions in other case studies.

The particular case study presented in this paper lead to the conclusion that ontology-based image annotations can lead to good retrieval performance in terms of recall and precision and good queriability if an appropriate modelling technique is used. However, it turned out that this involves high modelling effort since manual intervention in the translation process is required. It would be worth to investigate whether it is possible to combine $\MA_2$ and $\MA_5$ by using text mining tools not only to automatically map concepts but also relational information. Another important observation we made is that SNOMED CT is suitable for modelling image descriptions and - most important - that a rather small customisation, i.e. adding some role axioms in order to bypass cumbersome SNOMED CT constructs such as \emph{roleGroups}, can lead to great benefits in terms of queriability and comprehension.
\bibliography{swat4ls}

\begin{thebibliography}{1}

\bibitem{Baader2003}
F.~Baader, D.~Calvanese, D.~L. McGuinness, D.~Nardi, and P.~F. Patel-Schneider,
  editors.
\newblock {\em {The Description Logic Handbook: Theory, Implementation, and
  Applications}}.
\newblock Cambridge University Press, 2003.

\bibitem{Calvanese1998}
D.~Calvanese, G.~D. Giacomo, and M.~Lenzerini.
\newblock {On the Decidability of Query Containment under Constraints}.
\newblock In {\em PODS}, pages 149--158, 1998.

\bibitem{Chandler1976}
J.~S. Chandler and T.~G. DeLutis.
\newblock {A Methodology for the Performance Evaluation of Information Systems
  under Multiple Criteria}.
\newblock In {\em International Computer Measurement Group Conference}, pages
  221--229, 1976.

\bibitem{Harmon1999}
S.~Y. Harmon.
\newblock {Application of a Technique for Evaluating Information System
  Architectural Designs}.
\newblock In {\em {Symposium on Engineering of Computer-Based Systems}}. IEEE
  Computer Society, 1999.

\bibitem{Krauthammer2004}
M.~Krauthammer and G.~Nenadic.
\newblock {Term Identification in the Biomedical Literature}.
\newblock {\em Journal of Biomedical Informatics}, 37(6):512--526, 2004.

\bibitem{Morse2002}
E.~L. Morse.
\newblock {Evaluation Methodologies for Information Management Systems}.
\newblock {\em D-Lib Magazine}, 8(9), 2002.

\bibitem{owl-spec}
B.~Motik, P.~F. Patel-Schneider, and B.~Parsia.
\newblock {OWL 2 Web Ontology Language: Structural Specification and
  Functional-Style Syntax}.
\newblock Technical report, W3C Recommendation, 2009.

\bibitem{Opitz2009}
J.~Opitz, B.~Parsia, and U.~Sattler.
\newblock {Using Ontologies for Medical Image Retrieval - An Experiment}.
\newblock In {\em OWL Experiences and Directions (OWLED 2009)}, volume 529 of
  {\em CEUR Workshop Proceedings}. CEUR-WS.org, 2009.

\bibitem{Opitz2010}
J.~Opitz, B.~Parsia, and U.~Sattler.
\newblock {Information System Analysis}.
\newblock In {\em International Workshop on Evaluation of Semantic Technologies
  (IWEST 2010)}, CEUR Workshop Proceedings. CEUR-WS.org, 2010.

\end{thebibliography}
\bibliographystyle{abbrv}

\end{document}